\documentstyle[prl,aps,twocolumn,epsfig]{revtex}

\begin{document}

\preprint{}
\draft

\title{Stochastic Resonance and Nonlinear Response by NMR Spectroscopy}
\author{L. Viola${}^{1\,\dagger}$, E. M. Fortunato${}^2$,   
S. Lloyd${}^1$, C.-H. Tseng${}^2$, and D. G. Cory${}^2$ }
\address{ ${}^1$ d'Arbeloff Laboratory for Information Systems and 
Technology, 
Department of Mechanical Engineering, \\ 
Massachusetts Institute of 
Technology, 
Cambridge, Massachusetts 02139 \\
${}^2$ Department of Nuclear Engineering,    
Massachusetts Institute of Technology, 
Cambridge, Massachusetts 02139 }

\maketitle

\begin{abstract}
We revisit the phenomenon of quantum stochastic resonance in the regime
of validity of the Bloch equations. We find that a stochastic resonance 
behavior in the steady-state response of the system is present whenever 
the noise-induced relaxation dynamics can be characterized via a single 
relaxation time scale. The picture is validated by a simple nuclear 
magnetic resonance experiment in water. 
\end{abstract}

\pacs{03.65.-w, 05.30.-d, 76.60.-k}



The interplay between dissipation and coherent driving 
in the presence of dynamical nonlinearities gives rise to a variety of
intriguing behaviors. The most paradigmatic and counterintuitive example is 
the phenomenon of {\sl stochastic resonance} (SR), whereby the response of the 
system to the driving input signal attains a maximum at an optimum noise
level \cite{gammaitoni:1998}.  
By now, stochastic resonance has been demonstrated in several overdamped 
bistable systems as diverse as lasers, semiconductor devices, 
SQUID's, and sensory neurons, the required noise tuning being 
accomplished by either controlling the injection of external noise or by 
suitably varying the temperature of the noise-inducing environment.

Due to the broad typology of situations it can exemplify and its inherent 
simplicity, a preferred candidate for theoretical analysis is represented 
by a driven dissipative two-level system (TLS). The investigation has only 
recently been taken into the quantum world, where some prominent results have
been established for the so-called {\sl spin-boson model}. The latter 
schematizes the archetypal situation of a driven quantum-mechanical tunneling 
system in contact with a harmonic heat bath, the resulting dissipation being 
commonly addressed in the linear Ohmic regime 
\cite{lofstedt:1994,makri:1995,grifoni:1996,pareek:1997,fox:1998}. 
Within this framework, a quantum SR phenomenon 
induced by a resonant irradiation with a continuous-wave field 
has been characterized analytically \cite{makri:1995,pareek:1997} and verified 
through exact numerical path-integral calculations \cite{makri:1995,fox:1998}. 

In the present work, we show that a stochastic resonance phenomenon occurs 
for a much wider class of driven two-state quantum systems, whose relaxation 
dynamics can be accounted for by conventional Bloch equations. We find that, 
irrespective of the details of the microscopic picture, the essential 
requirement is the {\sl emergence of a single relaxation time scale}. 
The prediction is neatly demonstrated by a nuclear magnetic resonance 
experiment on a water sample.

Let us consider a two-state quantum system whose density operator 
$\rho$ is represented in terms of the Bloch vector $\vec{s}$ as 
$\rho=(1+\vec{s}\cdot \vec{\sigma})/2$ {\it i.e.},
$s_i(t)=\langle \sigma_i (t) \rangle$, $i=1,2,3$, in the customary 
pseudo-spin formalism \cite{allen:1975}.
Within the semigroup approach for open quantum systems \cite{alicki:1987}, 
the most general (completely) positive relaxation dynamics induced by the 
coupling to some environment is described by a quantum Markov master 
equation of the form
\begin{eqnarray}
\dot{\rho} = -{i \over \hbar} [H(t), \rho] + {1\over 2} 
\sum_{k,l=1}^3 a_{kl} 
\left\{ [ \sigma_k \rho, \sigma_l] + [ \sigma_k, \rho \sigma_l] \right\}\,. 
\label{qm}
\end{eqnarray}
The Hamiltonian $H(t)$, which describes the interaction of the TLS 
with the (classical) driving field, can be expressed as 
$H(t)= {\hbar \omega_0} \sigma_3/2 + V(t)$, $ V(t)= {\hbar (2 \omega_1)}
\cos (\Omega t) \sigma_1/2$,
where the Larmor frequency $\omega_0 =(E_2 - E_1)/\hbar$ associated with 
the TLS energy splitting and the Rabi frequency $\omega_1$
proportional to the alternating field amplitude have been introduced.  
The above Hamiltonian is identical to the one describing a driven
tunneling process in a symmetric double-well system with ``localized''
states provided by $\sigma_1$-eigenstates: By rotating the spin 
coordinate by $\pi/2$ about the $\hat{y}$-axis, one formally recovers 
the picture of tunneling in the $\hat{z}$-representation that is 
encountered in the literature 
\cite{gammaitoni:1998,lofstedt:1994,makri:1995,grifoni:1996,pareek:1997,fox:1998,leggett:1987}. 
The dissipative component of the TLS dynamics is fully 
characterized by the positive-definite $3\times 3$ relaxation matrix 
$A=\{a_{kl}\}$, determining the equilibrium state of the system
and the relaxation time scales connected with the equilibration 
process. The Bloch equations correspond to an especially simple 
realization of $A$, the non-zero elements being specified in terms of 
3 independent parameters: $a_{11}=a_{22}=(2T_1)^{-1}$,  $a_{33}= 
(T_2)^{-1}-(2T_1)^{-1}$, $a_{12}=a_{21}^\ast=i (\sqrt{2} T_1)^{-1} s_{eq}$.
$T_1, T_2$, and $s_{eq}$ are identified as the longitudinal and transverse 
lifetimes, and the equilibrium value of the population difference 
respectively. Thus, Eq. (\ref{qm}) takes the following familiar form 
\cite{bloch:1946}:
\begin{eqnarray}
\left\{ \begin{array}{lll}
\dot{s_1} & = & \;\; \: \omega_0 s_2  - T_2^{-1} s_1 \:, \\
\dot{s}_2 & = & -\omega_0 s_1 - T_2^{-1} s_2 + 2\omega_1 \cos( \Omega t) 
s_3 \:, \\
\dot{s}_3 & = &  -2\omega_1 \cos (\Omega t) s_2 - T_1^{-1}(s_3-s_{eq}) 
\:. \end{array} \right. 
\label{bloch}
\end{eqnarray}

In microscopic derivations of (\ref{bloch}), including the 
ones based on the spin-boson model in the appropriate limit 
\cite{leggett:1987,pareek:1997}, relaxation is caused  
by elementary processes involving noise-assisted transitions between 
the TLS energy levels or purely dephasing events with no energy 
exchange between the system and the environment. 
The overall relaxation rates $T_{1,2}^{-1}$ are
obtained by integrating such fluctuation and dissipation effects over the 
environmental modes, weighted by the appropriate noise spectral densities. 
Since the latter contain the coupling strength between the system and the 
environment, relaxation rates are themselves directly proportional to the 
underlying noise intensity.
Note that the above treatment in terms of a constant matrix $A$
is only valid for external fields that are relatively weak on the TLS 
energy scale {\it i.e.}, $2|\omega_1| \ll \omega_0$. In spite of the 
many restrictions involved, it is remarkable that the Bloch equations
(\ref{bloch}) are of such a wide applicability to cover the majority of 
magnetic or optical resonance experiments. 

For times long compared to the time scales $T_{1,2}$ of 
the transient dynamics, the motion of the system 
reaches a steady-state behavior that is insensitive to the initial condition
and acquires the periodicity of the driving. In particular, the asymptotic 
TLS coherence properties are captured by the off-diagonal matrix element
$s_1(t)$, $s_2(t)$. It is standard practice to formulate an input/output 
problem, where the TLS is regarded as a dynamical system generating 
$s_1(t)=\langle \sigma_1(t)\rangle$ as the output signal in response to a 
given input drive $V(t)$. 
By letting $\lim_{t\rightarrow \infty} \langle \sigma_1(t) \rangle =
s_1^{\infty}(t)$ denote the limiting steady-state value of $s_1(t)$, a 
figure of merit for the system response is the so-called {\sl fundamental
spectral amplitude} \cite{gammaitoni:1998},
\begin{equation}
\eta(\Omega,\omega_1)= |s_1^{\infty}(t)|= \hbar(2\omega_1) 
| \chi (\Omega,\omega_1)| \:,
\label{spectral}
\end{equation}
where the connection to the complex susceptibility 
$\chi (\Omega,\omega_1) = \chi' (\Omega,\omega_1)- i \chi'' (\Omega,\omega_1)$ 
is made explicit.

The competition between driving and dissipative forces 
sets the boundary between the linear vs. nonlinear response regimes. In the 
limit where $\omega_1 \ll T_{1,2}^{-1}$, only first-order contributions in 
$\omega_1$ are significant and the susceptibility $\chi(\Omega,\omega_1)
= \chi(\Omega)$ in (\ref{spectral}) can be calculated within ordinary linear
response theory. The linear regime implies that absorption of energy from the 
applied field occurs without disturbing populations from their equilibrium 
value $s_{eq}$. 
Linear behavior breaks down whenever $\omega_1 \gtrsim T_{1,2}^{-1}$. 
Strongly nonlinear-response regimes can be entered for arbitrarily weak fields 
as long as the coupling to the environment and the induced noise effects are 
weak enough. For both linear and nonlinear driving, the amplitude 
$\eta(\Omega,\omega_1)$ of the output signal also depends on the various 
parameters characterizing the noise process. 
Quite generally, the phenomenon of stochastic resonance can be associated 
with the appearance of {\sl non-monotonic dependencies upon  
noise parameters}, leading to the optimization of the response at a {\sl 
finite} noise level.

We focus on the Bloch equations (\ref{bloch}) with resonant driving, 
$\Omega = \omega_0$. It is then legitimate to invoke the rotating-wave 
approximation and replace the alternating field $V(t)$ with 
$V(t)= \hbar \omega_1 \cos (\Omega t) \sigma_1/2 - 
       \hbar \omega_1 \sin (\Omega t) \sigma_2/2$. The rotating-frame 
description of the Bloch vector $\vec{\mu}$ is introduced via the 
time-dependent rotation $R=\exp(i\Omega \sigma_3 t/2)$ {\it i.e.}, 
$\rho_R=R\, \rho \,R^{-1}=(1+\vec{\mu}\cdot\vec{\sigma})/2$, 
$\vec{\mu}=(u,v,w)$ \cite{allen:1975}. 
The steady-state solution to the Bloch equations is well known 
\cite{torrey:1949,abragam:1961}. In particular, 
$\chi'$ and $\chi''$ are read from the dispersive and absorptive 
components $u,v$ of the Bloch vector respectively, and the spectral amplitude
$\eta(\Omega=\omega_0,\omega_1)= (u^2+v^2)^{1/2}$. A simple expression is
found for the nonlinear response:
\begin{equation}
\eta(\Omega=\omega_0,\omega_1)=
s_{eq} \,\frac{\omega_1 T_2} {1+ \omega_1^2 T_1 T_2}\:. 
\label{response}
\end{equation}

Suppose now that we have the capability of manipulating the strength of the 
coupling of the TLS to its environment, thereby changing the relaxation 
times $T_1,T_2$. For a fixed driving amplitude $\omega_1$, $\eta$ displays 
purely monotonic behaviors if $T_1,T_2$ are varied independently. 
However, if a {\sl single} relaxation time is present, 
$T_1 = T_2 =T_{12}$, $\eta$ develops a local maximum characterized by
\begin{equation}
T_{12}^\ast = \omega_1^{-1}\:, \hspace{5mm} \eta(\Omega=\omega_0, 
\omega_1,T_{12}^\ast)= {s_{eq} \over 2} \:. 
\label{sr}
\end{equation}
The occurrence of such a peak in the steady-state response as a function of 
the noise strength can be pictured as a stochastic resonance effect in the 
TLS. Physically, the condition for the maximum in (\ref{sr}) can be thought 
of as a synchronization between the periodicity of the rotating-frame vector 
in the absence of relaxation and the additional time scale emerging when 
dissipation is present. In semiclassical terms, a constraint of 
the form $T_1=k T_2$, $k=$ const., indicates that the spectral densities 
of the fluctuating environmental fields along different directions are 
{\sl not} independent upon each other. Simple examples include noise 
processes that effectively originate from a single direction or that 
equally affect the system in the three directions. In fact, the 
existence of a {\sl single} relaxation time is a feature shared 
with earlier investigations of quantum SR based on the driven spin-boson model
\cite{makri:1995,pareek:1997}, where it arises as a necessary consequence 
of the initial assumption that environmental forces exclusively act along 
the tunneling axis. However, we emphasize that our discussion is done 
{\sl without} reference to a specific model, encompassing in principle 
a larger variety of physical situations.

Apart from this conceptual difference, the SR phenomenon evidenced 
above is characterized by the same distinctive features found for  
the spin-boson model on resonance \cite{makri:1995,pareek:1997}.
According to (\ref{sr}), the maximum steady-state response is independent
of the driving amplitude, whereas the position of the peak shifts
toward shorter relaxation times with increasing $\omega_1$. Thus,
weaker noise strengths require weaker input fields to attain a large 
response. This brings the nonlinear nature of the SR mechanism to light,
for weaker dissipation more easily pushes the system into a regime where 
$\omega_1 T_{12} \gtrsim 1$. No SR peak occurs in the limit $\omega_1 T_{12} 
\ll 1$ where linear response theory applies and $\eta \rightarrow 
s_{eq}(\omega_1 T_{12}) \ll s_{eq}$. 
Thus, SR results in efficient noise-assisted signal amplification. 
Breakdown of linear behavior is more convincingly demonstrated by 
looking at the dependence of the response (\ref{response}) upon the 
external field strength. It is easily checked that the condition 
(\ref{sr}) simultaneously optimizes the 
response against $\omega_1$, with $\omega_1^\ast=(T_1 T_2)^{-1/2}$.
However, {\sl it is only when $T_1=k T_2$ that the existence of such an 
optimal field amplitude coexists with a SR effect.}    
 
Nuclear Magnetic Resonance (NMR) provides a natural candidate for a 
direct experimental verification of the predicted phenomenon. In NMR, 
the Bloch equations (\ref{bloch}) describe the motion of the magnetization 
vector $\vec{M}$ of spin 1/2 nuclei ($^1$H) that are subjected to a 
static magnetic field $B_0$ along the $\hat{z}$-axis and a radio-frequency 
signal with amplitude $2 |B_1| \ll B_0$ applied at frequency $\Omega$ along 
the $\hat{x}$-axis. The mapping is established by identifying 
$\vec{s}=\vec{M}$, $s_{eq}=M_0$, $\omega_0=\gamma B_0$, 
$\omega_1= \gamma B_1$, $M_0$ and $\gamma$ denoting the equilibrium 
magnetization and the gyromagnetic ratio respectively. 
Relaxation processes arise due to a multiplicity of microscopic mechanisms 
\cite{abragam:1961}. 
For a liquid spin $1/2$ sample, the leading contribution arises from 
fluctuations of the local dipolar field caused by bodily motion of 
the nuclei.
The longitudinal relaxation time $T_1$ is essentially determined by 
the $\hat{x}$- and $\hat{y}$- components of the local magnetic fields 
at the Larmor frequency, while the transverse lifetime $T_2$ takes extra 
contributions from static components of the $\hat{z}$-field, implying 
that $T_2 \leq 2 T_1$ ordinarily \cite{abragam:1961,slichter}.
Let us assume as above that $\Omega=\omega_0$. 
Once the steady state is reached, 
the magnetization vector $\vec{M}$ precesses about the $\hat{z}$-axis with the 
periodicity of the r.f. field. The variation of the dipole moment in the 
tranverse plane induces a measurable e.m.f. in a Faraday coil. This
provides access to the relevant quantity $\eta$ of Eqs. 
(\ref{spectral})-(\ref{response}), which represents the length of the 
transverse magnetization vector, $(u^2+v^2)^{1/2}=(M_x^2(t)+M_y^2(t))^{1/2}$.

Our experiment consists in probing the steady-state magnetization 
response of water {\sl as a function of the noise strength inducing the 
natural relaxation processes}. The $^1$H Larmor frequency at $B_0= 9.4$ T 
is $\omega_0/2\pi= 400$ MHz, with relaxation times $T_1=3.6$ s, $T_2=2.5$ s. 
The sample can be brought to a regime where $T_1\simeq T_2$ upon
addition of the paramagnetic salt copper sulfate (CuSO$_4$). 
With concentrations in the range between 40 mM and 100 mM, collision
events with the impurity dominate the nuclear relaxation dynamics. This 
effectively pushes the system into a regime of {\sl rapid motion} where the 
correlation time of the local magnetic fields seen by the nuclei is very 
short on the scale $\omega_0^{-1}$, thereby ensuring that 
$T_1=T_2=T_{12}$ \cite{slichter}. Higher concentrations of the 
CuSO$_4$ additive result in a shorter relaxation time 
$T_{12}$ hence implying an effective tuning of the noise strength.  
All measurements were performed at room temperature with a 
Bruker AMX400 spectrometer on five water samples with additive 
concentration in the above range. 

Independent measurements of $T_1$ and $T_2$ were made to confirm that the 
amount of CuSO$_4$ was sufficient to make them equal.
$T_1$ was measured via an inversion recovery technique
\cite{slichter}, by looking at the recovery curve of $M_z(t)$ after
the application of a $\pi$ pulse causing $M_z(0)= - M_0$. 
Values of $T_2$ were inferred from the decay of the echo signals in a
standard Carr-Purcell sequence where $\pi$ rotations were used 
to refocus dephasing due to inhomogeneous broadening \cite{slichter}.
For the 5 concentrations utilized, $T_1$ and $T_2$ were found to 
be within 1 ms of the average value $T_{12}$ which is listed for each 
sample in Table I. 
\begin{center}
\begin{tabular}{|c|c|} \hline
	CuSO$_4$ (mM) & $T_{12}$ (ms) $\pm$ 1 ms \\ \hline\hline
		40		        &	45.5 \\
		50			&	36.5 \\
		60			&	28.5 \\
		75			&	25.0 \\
		100			&	18.0 \\ \hline\hline
\end{tabular}
\end{center}
\begin{flushleft}
{\small TABLE I. Relaxation time $T_{12}=T_1=T_2$ as a function of the
CuSO$_4$ paramagnetic impurity concentration for the 5 water samples 
used in the experiment.} 
\end{flushleft}

For each sample, the response to a long external r.f. pulse was measured 
for various values of the driving amplitude. For a given driving
amplitude, the duration of the pulse was increased up to about 200 ms and 
the reading was continued until a constant e.m.f. value was reached, 
confirming that all transient responses had sufficiently decayed. 
Under these conditions, the observed steady-state value is equivalent to 
the one produced by a cw-irradiation as assumed in Eqs. (\ref{bloch}). 
A delay long with respect to $T_{12}$ was waited between each pulse to 
allow the sample to return to equilibrium. The value of the r.f. amplitude
was determined by extrapolating measurements of the nutation rate 
$\nu_1=\omega_1/2\pi$ at a high field setting down to the 
relevant lower-field domain in the neighbourhood of 
$\omega_1 \approx T_{12}^{-1}$. 
While the relative error between two r.f. setting is found to be small, 
the systematic error associated with the extrapolation turns out to be
significant. A linear correction of the frequency scale was included in the 
analysis to compensate for such error.

The experimental results are shown in Figs. 1 and 2. Fig. 1 
evidences the SR peak for three values of the driving amplitude. 
A bell-shaped maximum in the response profile is clearly visible, as well as
the expected shifting of the peak location with increasing $\omega_1$. For
each curve, the SR condition $T_{12}^\ast \approx \omega_1^{-1}$ of Eq.
(\ref{sr}) is in fairly good agreement with the observed behavior, existing
discrepancies being accounted for by the residual error 
affecting the determination of $\omega_1$. In Fig. 2 the complementary
characterization of the SR effect in terms of nonlinear response to the 
driving field is displayed for three values of $T_{12}$. In each case,
ordinary linear response theory is valid for small $\omega_1$ to the left 
side of the maximum. Thus, SR reveals itself as a signal optimization marking 
the crossover between linear and nonlinear response. 

Beside validating the predictions from the Bloch equations (\ref{bloch}),
our experimental results also support the conclusions 
independently reached in earlier theoretical analyses 
\cite{makri:1995,pareek:1997}. 
A few remarks are in order concerning the specific case of NMR.
First, the present experiment is {\sl not} a stochastic NMR experiment
\cite{snmr}. While the obvious similarity is that both methods 
probe the nuclear spin system by looking at the transverse magnetization 
response, in stochastic NMR the system is directly excited by noise, 
which is therefore always {\sl extrinsic} (and classical) in origin. 
More importantly, as mentioned already, the solutions to the Bloch 
equation have a long history as a tool to investigate magnetic 
resonance behaviors. In particular, the {\sl existence} of an optimum 
r.f. amplitude $\omega_1^\ast= (T_1 T_2)^{-1/2}$ 
is a feature pointed out long ago by Bloch himself 
\cite{bloch:1946}. However, only the SR paradigm provides the motivation 
to regard relaxation features as controllable output parameters and to 
look at the usual response behavior {\sl along different axes in the 
parameter space}. 
Even once this is done, this does not automatically lead to SR. 
Rather, it is the recognition that a {\sl single} axis $T_1=T_2$ is 
effectively needed to bring out the fingerprint of the phenomenon 
and the novel element added to the standard NMR analysis.

In summary, we established both theoretically and experimentally
the occurrence of stochastic resonance in two-state quantum systems
whose relaxation dynamics are described by Bloch equations. In addition to
substantially broadening the existing paradigm for stochastic resonance 
in quantum systems, our results point to the possibility 
of characterizing intrinsic relaxation behavior via resonance effects.
By offering an optimized way for input/output transmission against  
noise, stochastic resonance carries a great potential for 
systems configured to perform specific signal processing and communication 
tasks \cite{gammaitoni:1998}. In particular, full exploitation of  
stochastic resonance phenomena could potentially 
disclose a useful scenario for reliable transmission of quantum 
information in the presence of environmental noise and decoherence. 

This work was supported by the U.S. Army Research Office under grant number
DAAG 55-97-1-0342 from the DARPA Microsystems Technology Office.

${}^\dagger$ Corresponding author: {\tt vlorenza@mit.edu}


\begin{figure}[h]
\epsfxsize=8cm
\mbox{\epsfbox{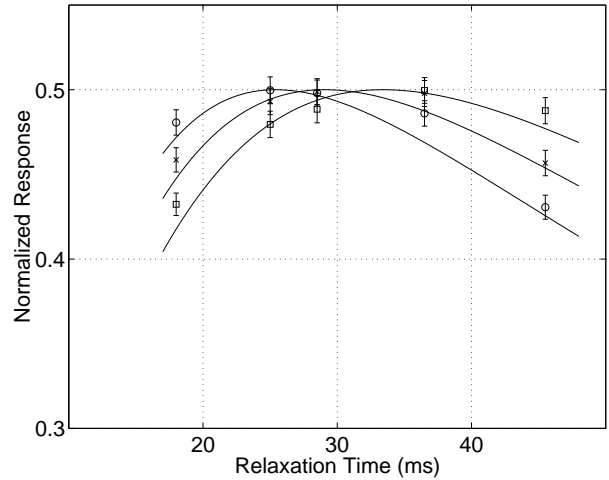}}
\vskip .2cm
\caption{ Normalized steady-state response $\eta/s_{eq}$ vs. relaxation 
time $T_{12}$ for resonant driving $\Omega/2\pi =400$ MHz at different 
driving amplitudes: $\omega_1/2\pi$= 6.3 Hz (circles), 5.5 Hz (crosses),
4.8 Hz (squares). Solid lines: theoretical predictions from Eq. (4), after 
systematic correction for the frequency (no free parameters). }  
\label{figure1}
\end{figure}

\begin{figure}[h]
\epsfxsize=8cm
\mbox{\epsfbox{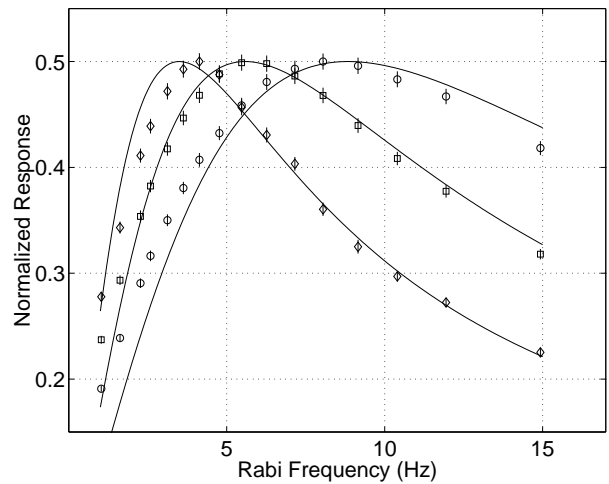}}
\vskip .2cm
\caption{ Normalized steady-state response $\eta/s_{eq}$ vs. Rabi
frequency $\omega_1/2\pi$ for resonant driving $\Omega/2 \pi=400$ MHz at 
different relaxation times: 
$T_{12}$= 18.0 ms (circles), 28.5 ms (squares), 45.5 ms (diamonds).
Solid lines as above. }   
\label{figure2}
\end{figure}

\end{document}